\documentclass[runningheads]{llncs}
\usepackage{changepage}
\usepackage{graphicx}
\usepackage[colorinlistoftodos]{todonotes}
\usepackage{microtype}
\usepackage{xspace}
\usepackage[newcommands]{ragged2e}

\begin{document}
\title{Building a Legal Dialogue System: Development Process,\\ Challenges and Opportunities}
\titlerunning{Building a Legal Dialogue System}
\author{Mudita Sharma\inst{1} \and
Tony Russell-Rose\inst{2} \and
Lina Barakat\inst{3} \and
Akitaka Matsuo\inst{4}}
%
%
\institute{University of Essex \\\email{mudita.sharma19@gmail.com} \\\and
\email{tr18907@essex.ac.uk}\\\and
\email{lina.barakat@essex.ac.uk}\\\and
\email{a.matsuo@essex.ac.uk}
}
\maketitle              
\newcommand{\faq}{\textsc{FAQ}\xspace}
\newcommand{\ff}{\textsc{FF}\xspace}
\newcommand{\lex}{\textsc{lex}\xspace}
\newcommand{\aws}{\textsc{aws}\xspace}
\newcommand{\cli}{\textsc{cli}\xspace}

\begin{abstract}
This paper presents key principles and solutions to the challenges faced in designing a domain-specific conversational agent for the legal domain. It includes issues of scope, platform, architecture and preparation of input data. It provides functionality in answering user queries and recording user information including contact details and case-related information. It utilises deep learning technology built upon Amazon Web Services (\aws) \lex in combination with \aws Lambda. Due to lack of publicly available data, we identified two methods including crowdsourcing experiments and archived enquiries to develop a number of linguistic resources. This includes a training dataset, set of predetermined responses for the conversational agent, a set of regression test cases and a further conversation test set. We propose a hierarchical bot structure that facilitates multi-level delegation and report model accuracy on the regression test set. Additionally, we highlight features that are added to the bot to improve the conversation flow and overall user experience.


\end{abstract}

\section{Introduction}
In recent years conversational agents have become efficient means for supporting business across a variety of industry sectors. Chatbots and other dialog systems are now routinely deployed for various tasks such as booking appointments, answering users' queries, generating leads and recommending products and services to the user. Consequently, a variety of cloud based and non-cloud based platforms have been launched for their development and hosting such as 
\aws \lex~\footnote{\url{https://aws.amazon.com/lex/}}, Google's Dialogflow~\footnote{\url{https://cloud.google.com/dialogflow}} and Rasa~\footnote{\url{https://rasa.com/}}. 
These platforms have revolutionized the process of building a dialog system and the research on which they are based has led to significant improvements to the quality of the conversational user experience. 

Dialog systems can be classified in three primary categories~\cite{gao2018neural}: question answering systems that answer specific user queries, task-oriented systems that assist with specific tasks such as booking appointments, and social bots whose purpose is to make human-like conversation and act more like a companion to their users. 

Traditional dialog systems involve a complex pipeline whose components interact to offer a human-like conversation~\cite{zue2000juplter}.
From an architectural perspective, social bots may use neural networks such as encoder-decoder or seq2seq models~\cite{serban2015building,vinyals2015neural,bordes2016learning} and other generative models to generate responses to user input. By contrast, retrieval-based systems map an input to a response from a repository of responses. A recent survey on dialog systems can be found in~\cite{chen2017survey,chen2018deep,serban2015survey}. 

At the outset we chose to build a retrieval-based system rather than a generative system. The former approach is popular in commercial applications due to need for less resources, cost effective and easy maintenance. This paper focuses on the architecture, linguistic resources and additional design features involved in building a dialog system specifically for the legal sector, although it is our belief that many of the insights are in fact agnostic of domain.

We present a composite system that combines question answering and task-oriented approaches, focusing on two specific tasks discussed in section~\ref{step by step}. The rest of the paper is organised as follows. We also explain the bot hierarchical structure. Section~\ref{sec:architecture} describes different components of the end-to-end prototype system. This is followed by section~\ref{sec:ling-res} on the linguistic resource describing the legal dataset. Section~\ref{sec:add features} is an examination of the key challenges we faced and the additional features we included to improve the conversational flow. In section~\ref{conclusion} we conclude the paper with a summary and future work.

\section{Approach and Development of the Conversational agent}\label{step by step}
We present the approach we followed in identifying requirements and selecting a suitable platform to build the conversational agent. This is followed by the description of the bot hierarchical design structure.

\subsection{Identifying requirements}\label{sec:Identifying requirements}
The first step in building a dialog system for the legal domain is to understand the key processes and functions that take place within a law firm. This involves analysis of user journeys and enquiry data from case management systems to identify use cases and tasks that a dialog system could support. In this paper, we focus on two key use cases,
\begin{enumerate}
    \item Frequently asked questions (\faq): answering general questions related to the services offered by a law firm (note that this is not the same as giving legal advice specific to user). Queries such as this are typically fulfilled in a single exchange, i.e. by providing a single answer to a user's question, for example:\\
    \textit{User}: Can I bring my partner to the appointment?\\
    \textit{Sys}: Yes you can bring your partner to the appointment.\\
    
    \item Fact finding (\ff): The system attempts to identify the particular service the user needs based on their specific case description then records their contact details and the case description. 
    This use case is fulfilled by recording the details and usually involves a multi-turn conversation, for example:\\
    \textit{User}: I want someone to review my contract.\\
    \textit{Sys}: Sure, to help you with that we would need your contact details for someone from the firm to contact you.\\
    \textit{Sys}: What is your name?\\
    \textit{User}: Jon \\
    \textit{Sys}: What is your phone number?\\
    \textit{User}: 07423333333\\
    \textit{Sys}: Perhaps you can describe the type of contract you want reviewed.\\
    \textit{User}: A housing contract.\\
    \textit{Sys}: Thanks for that. One of our legal experts will contact you as soon as possible.
\end{enumerate}
The system currently supports \faq and \ff for various legal services, including:
\begin{itemize}
  \item ten services from commercial corporate
  \item five from wills and life planning
  \item three from commercial property
  \item one from civil dispute resolution
  \item one from immigration
\end{itemize}
Example services include Contract Review (\textit{CR}) and Draft and Update Terms and Conditions (\textit{DUTnC}) from commercial corporate. 

The dialog model is based on the notion of identifying \emph{slots} from the user text and user \emph{intents}, which are defined according to the specific use case.
A slot can have different related values in the input, for example the \textit{``day"} slot can have values according to days in the week. A slot can be predefined with a set of accepted slot values. 
We define multiple slots to be identified in the user utterance, including consent to enter details that takes "yes" or "no" as input, location of the firm, etc. 
To identify different legal services within a given user input, we define a custom slot ``\textit{practice\_type}" with various legal services as its slot values. 

An intent is the user's objective behind their utterance. For example, user text ``What is the firm's location?" implies user's intent to retrieve a list of locations of the firm. 
For the FAQ use case, we define intents according to the input type. For example, a ``\textit{Cost}" intent is used to deal with questions relating to the price of different legal services, while a ``\textit{Prep\_app}" intent handles questions related to the documents required to bring to an appointment. Each of these \faq intents will elicit a different response from the response resource (shown in Figure~\ref{fig:software design} and described in section~\ref{sec:architecture}) depending on the identified service in the slot.
By contrast, \ff intents are defined according to 
the specific legal service description. For example ``I want someone to look at a contract for me." and ``I need help with drafting a contract." may look similar but they refer to different services (``\textit{CR}") and (``\textit{DUTnC}") respectively.
Our initial attempt to conflate these intents led to confusing responses, as it is difficult for the model to accurately identify the service from the description of a legal case such as in the examples above. It is therefore vital in \ff to use the whole utterance as a description of the user's legal case to identify and populate legal service slots. 
Consequently, \ff intents are defined on a per-service basis whereas in \faq intents are 
based on the combination of input type and particular service. 

\subsection{Selecting a Platform}\label{subsec:selecting a platform}
The next step is to select a platform from those available such as Google's Dialogflow, Rasa and \aws \lex that provides appropriate functionality to develop and deploy the dialog system. 

After analysing various frameworks, we selected \lex for its simple development environment, comprehensive documentation, rapid development cycle, use of deep learning and other advanced natural language understanding functionality for intent recognition. It is a service offered by Amazon Web Services (\aws) that provides an easy-to-use IDE to define intents and entities.
\lex also supports speech input and output, although at present our use of the platform extends only to text based dialogues. 

In addition to defining intents specific to two use cases, we use few intents predefined by \lex. \lex provides a variety of inbuilt intents such as ``\textit{greet\_intent}" and ``\textit{goodbye\_intent}". We use the former to identify utterances such as ``hi" and ``good morning" and the latter for utterances such as ``bye" and ``goodbye".
\lex also provides inbuilt slots such as ``\textit{First\_name}", ``\textit{Last\_name}", ``\textit{Number}" and ``\textit{Email\_address}" which we use to persist user details in \ff. 

\subsection{Bot Hierarchy}
\begin{figure*}
    \centering
    \includegraphics[scale = 0.59]{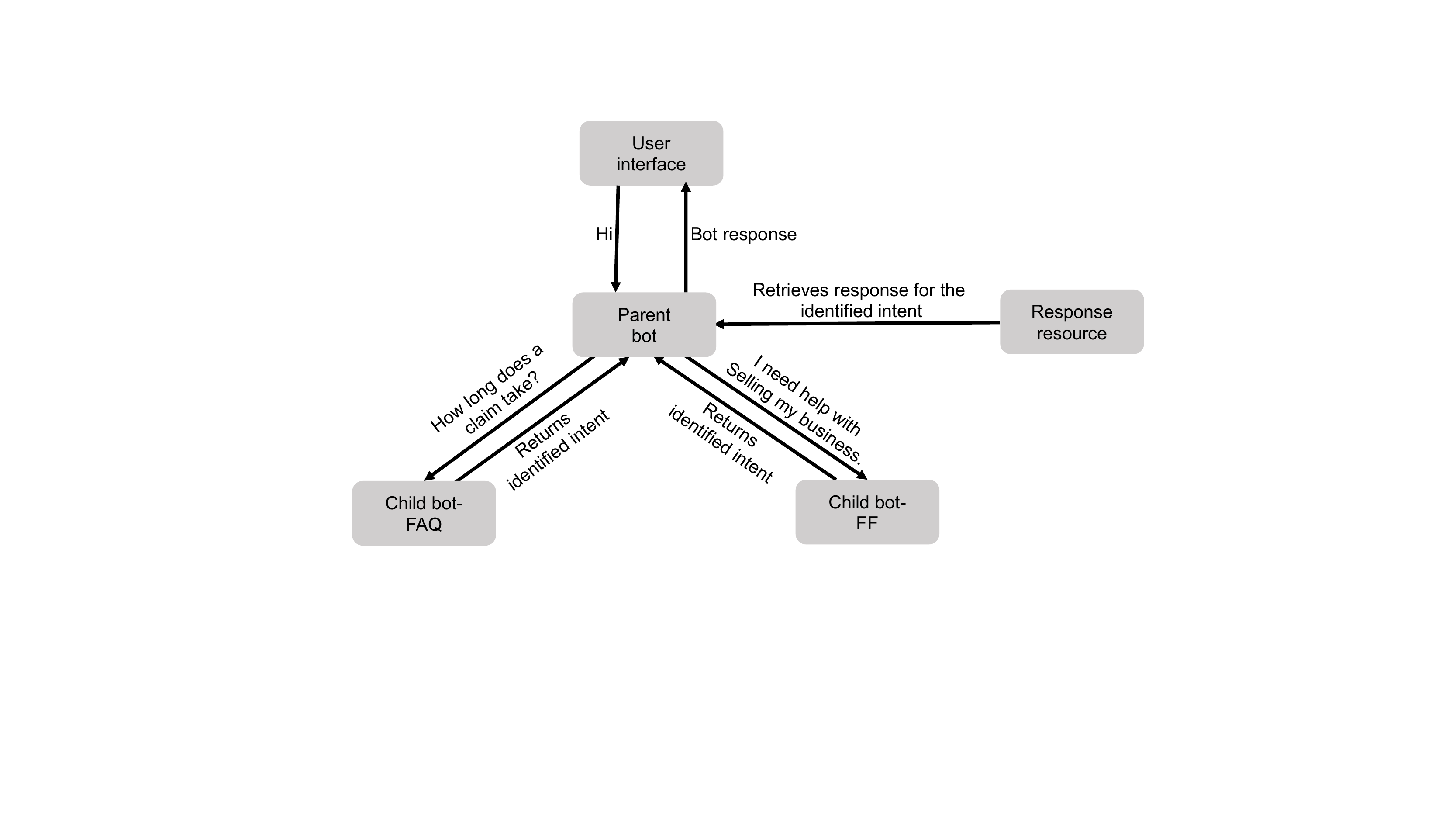}
    \caption{Bot hierarchy}
    \label{fig:Bot hierarchy}
\end{figure*}
%
We use a hierarchical structure of multiple bots to overcome the limits on the number of permitted intents and slots within \lex, but for the structured view of intent distribution in different use cases and to make the maintenance and improvement easier under such complicated intent relations.
Three unique bots structured at two levels are designed to share a parent-child relationship as shown in Figure~\ref{fig:Bot hierarchy}. The intents are divided amongst three bots such that the parent bot consists of $14$ intents and child bots with a total of $44$ intents (child bot-FF with $19$ intents and child bot-FAQ with $25$ intents). 

As shown in the figure, user interacts directly with the parent bot via user interface. The parent bot handles the basic user intents such as ``\textit{greet\_intent}" and ``\textit{goodbye\_intent}". In addition to these intents, we include two additional custom intents in the parent bot. A custom intent, \textit{all\_faq}, represents sentences from all intents of a child bot-\faq and the other intent, \textit{all\_ff}, consists of all sentences defined in the second child bot child bot-\ff. Thus, when the parent bot receives \ff or \faq utterance, the underlying model maps (or classifies) the utterance as either \faq or \ff query based on the \textit{all\_faq} or \textit{all\_ff} intent invoked. The query is then passed to the meta-classifier for the respective child bot that identify and return a specific intent recognised for the input utterance. We define child bot-\faq and child bot-\ff that receives and classifies \faq and \ff related queries respectively. Once the parent bot receives an intent for the input utterance from one of the child bots, it returns the corresponding response to the user. The response returned by the system is deterministic. This means that once the parent bot receives intent from a child bot, a response is retrieved from a predefined dataset known as the response resource described in section~\ref{sec:architecture}.

The bot hierarchy can be further extended to include multiple levels of delegation. For instance, to include more use cases, multiple child bots at the same level or at different levels can be integrated such that bots at the same level are independent from each other (i.e. include intents with mutually exclusive utterances) and the connected bots at different levels communicate with each other. That is, the bots at same level cannot facilitate delegation whereas the bots at different levels can. Each bot, irrespective of level, can be designed to have a unique classifier.

\section{Architecture}\label{sec:architecture}
\begin{figure*}
    \centering
    \includegraphics[scale = 0.59]{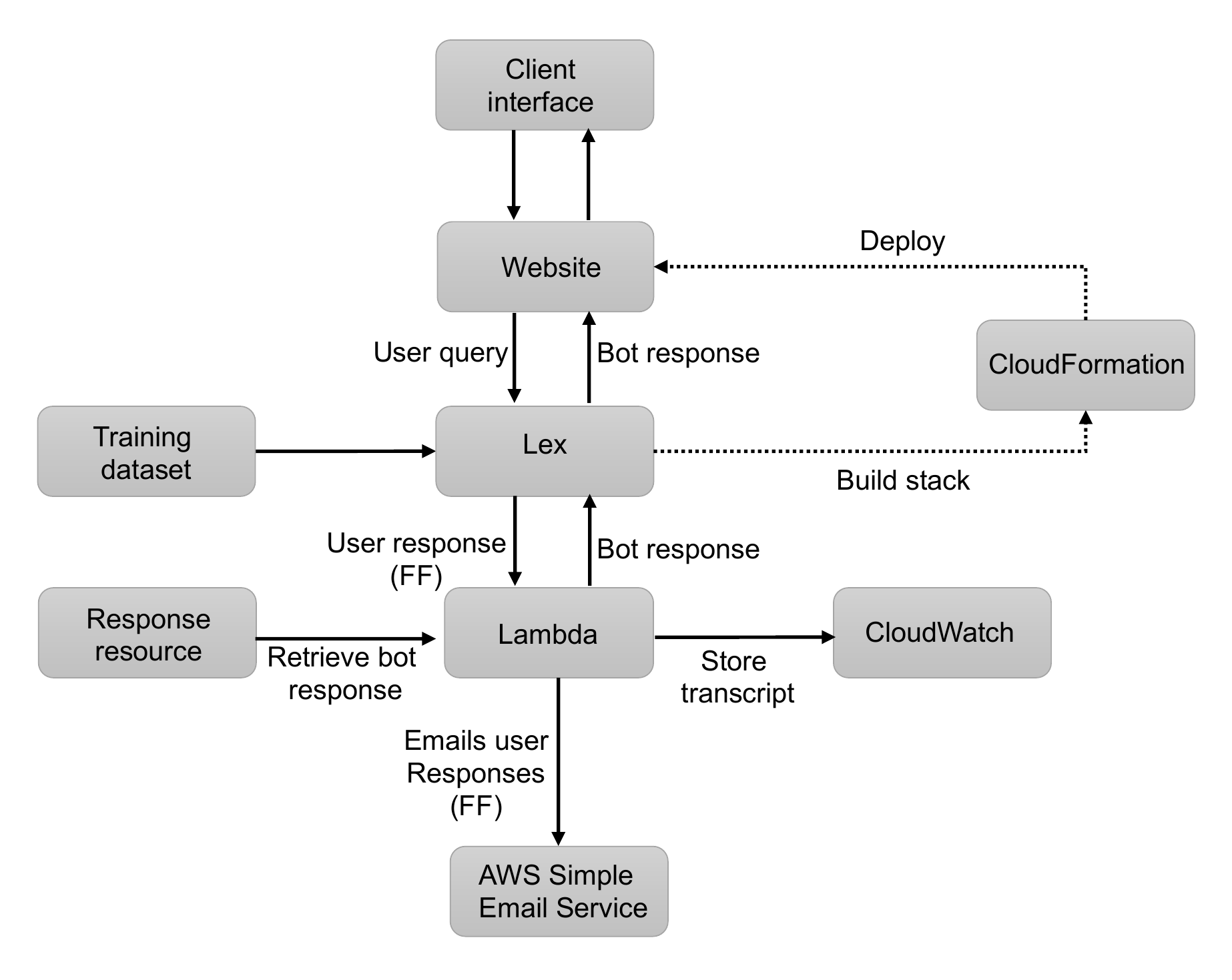}
    \caption{Software architecture}
    \label{fig:software design}
\end{figure*}

A bot in a commercial setting includes various building blocks from development to deployment as shown in Figure~\ref{fig:software design}. We discuss in detail each component in the prototype development of a conversational agent. 

An important component of this architecture is \lex, an AWS chatbot service described in the previous section~\ref{subsec:selecting a platform}. \lex has a chat window to facilitate conversational dialogue between the agent and the user. The user chats with the agent via chat box. The chat box is present on a chat interface such as a website or as an app on a collaborative platform such as slack. The chat is purely text-based and includes graphical features such as buttons to display different slot choices. For the end-to-end service, we utilise other services provided by \aws which are described further in this section.

We utilise \aws serverless computing platform, \aws Lambda~\footnote{\url{https://aws.amazon.com/lambda/}} in the backend for various purposes. First of all, a Lambda function consists of algorithms to define and control the conversation flow between the user and \lex. We chose to use Python, among various programming languages by \aws Lambda. 
A lambda function also talks with CloudWatch~\footnote{\url{https://aws.amazon.com/cloudwatch/}} to automate the process of monitoring conversations, store log of transcripts, troubleshoot issues and visualise logs.
In the case of \ff usecase, another lambda function triggers \aws Amazon Simple Email Service (SES)~\footnote{\url{https://aws.amazon.com/ses/}} to send emails containing user responses. Thus, when the user has entered the contact details and case related information, these details are automatically sent to the inbox specifically created to receive emails from the bot. The inbox is accessible by key members at the firm such as IT team, marketing team and a set of solicitors. The user's legal information is used by the legal expert to understand the user’s legal requirements prior to engaging directly with the user.
Finally, to deploy the conversational agent on a website, we use \aws CloudFormation~\footnote{\url{https://aws.amazon.com/cloudformation/}}. It deploys a collection of \aws resources and dependencies to launch and configure them together as a stack.

In retrieval-based dialog models, the response returned by the system is deterministic. This means that once an intent is identified, a response is generated from a predefined dataset known as the response resource (see Figure~\ref{fig:software design}).
In our case, the response dataset consists of a set of responses per intent per service for \faq and per intent for \ff. These responses are acquired and curated with the help of legal experts.

\section{Linguistic resources}\label{sec:ling-res}
In the following sections we describe the process of acquiring the linguistic resources to train the deep learning model driving \lex.

\subsection{Data collection}
To improve the accuracy of intent recognition, a dataset is required for training the dialog model that consists of various paraphrases for each intent. One of the challenges in building a legal dialog system is the lack of publicly available datasets (possibly due to confidentiality issues). 
To address this, we identified three sources of data which are described in this section in turn. 
An initial training dataset was generated by subject matter experts and representatives from each service such as solicitors, secretaries and paralegals. We refer to this henceforth as the 'baseline dataset' used to build and train the initial baseline model. We trained the baseline model on a total of $150$ utterances. Table~\ref{tab:sentence distribution} shows the number of utterances allocated to each use case. 

To improve the quality and quantity of the training data we investigated the use of crowdsourcing as a second data source. \cite{yu2016chatbot,bessho2012dialog} use a crowdsourcing approach to elicit diverse conversations and thus expand their dataset. Although their objective is similar to ours, there are two major differences. First, they invite participants in real-time on Twitter and Amazon Mechanical Turk with tests that are open to general public. By contrast, we deploy our baseline model and elicit responses via the Slack collaborative platform. Second, we recruit from a cohort of law students at a UK university as we require domain expertise and knowledge of legal terminology to elicit meaningful responses.
For example, the word ``will" can be used as a modal verb to express beliefs about the future, or used to refer to a document that describes distribution of assets. 
We recruited four law students to interact with the agent for a session of roughly $1.5$ hours. Each participant was assigned a private channel in the slack workspace and was given a number of hypothetical scenarios for various legal services. One of the scenarios is described below, \\

\begin{adjustwidth*}{0.15\linewidth}{}
Name: David Clark\\
Phone number: $+44123456$ \\
Email address: xyz279@abcd.com \\
David owns a telecommunication company, Telecom Corp. Ltd. There are $40000$ employees in his company. He needs help with drafting employment policies and procedures and would like information about some or all of the following: \\
length of whole process, visit Clacton office due to $<$some$>$ reason, length of the appointment, bring $<$someone$>$ to the meeting, attend meeting with $<$someone$>$, price, home visit, legal aid, prepare for an appointment, opening hours / days.\\
Finally, he wants to arrange an appointment with a solicitor.\\
\end{adjustwidth*}

In the scenario above, $<$text$>$ can be replaced with whatever the participant thinks is suitable. This brings diversity in the text. For example, visit Clacton office due to $<$some$>$ reason can constitute a question like ‘can I visit Clacton office as that is closest to my office?’. \\

The third data source constitutes the last three years' daily user enquiries received via an enquiry form located on the firm's website. These enquiries are not only a source of sentence paraphrases but also help identify new user intents which were later added to the dataset. For instance, some of the new intents identified are location of a legal firm, urgency of the legal matter and method of contact. The process of obtaining the data from third data source is described as follows. The data from this source is unstructured, that is the enquiry form usually consists of sender's name, subject, legal service, contact details, message body and other metadata. We have automated the process of extracting the message body from the enquiry. The rest of the content from the enquiry is discarded. Basically, we write a Python script that creates an excel file and writes the extracted message in the file. The script creates multiple sheets in the file each representing a legal service that is each sheet consists of enquiries related to a particular service.
Once all the enquiries are processed, we manually extract the most relevant utterances which becomes part of either training or test set as described further in this section. 

\subsection{Training and Test Data preparation}
The collected utterances from last two data sources are manually evaluated for relevance, appropriateness, and clarity. First, utterances that are not relevant to a specific context in the conversation, such as "I will keep an eye on my emails", are discarded. Second, the utterances are evaluated for appropriateness. For example, ''can you help me with employment law?" is a legitimate question, but out of scope as the dialog system is currently not trained to handle  enquiries related to employment law. Third, utterances are evaluated for clarity. For example, an utterance such as ''will validity" is ambiguous and therefore lacks clarity. Utterances that fail to meet these three criteria are discarded. The remaining utterances are then manually labelled with corresponding intents.
Finally, the labelled data is used as input to the system trained on the baseline dataset. Utterances that are correctly recognised are added to a set of regression test cases and remainder are added to the training dataset.
The regression test cases are used to ensure that the modifications to the system do not have a negative impact on the performance of the model. In future, we expect the training dataset and test cases to grow in size, and regression testing is performed every time the dialog model is modified or (re)trained. During regression testing, when an unseen utterance is encountered, \lex uses a classification technique to assign a confidence score to each intent based on intent classification accuracy, and the intent with the highest score is returned \footnote{\url{https://docs.aws.amazon.com/lex/latest/dg/confidence-scores.html}}.

\begin{table}[t]
\centering
\caption{Number of \faq and \ff sentences in training dataset}\label{tab:sentence distribution}
\begin{tabular}{c|r|r|r}
\hline
Data source & \faq & \ff & Total\\ \hline
\hline
Legal experts (Baseline) & 54 & 96 & 150\\ \hline
Crowdsource & 159 & 1 & 160\\\hline
\hline
\end{tabular}
\end{table}

%
%

$258$ utterances were collected from our crowdsourcing exercise. Out of these, we discarded $20$ utterances as they did not satisfy the three criteria described above. Of the remaining $238$ utterances, $78$ utterances were correctly recognised by the system and thus were added to the set of regression test cases. The remaining $160$ were added to the training dataset. As seen in  table~\ref{tab:sentence distribution}, crowdsourcing elicited only one further instance for the \ff use case (in addition to the $96$ in the baseline dataset). This is due to the fact that participants were able to choose a service using user interface buttons rather than textual input. Consequently, most crowdsourced utterances are for \faq rather than \ff.
In future, we plan to suppress these buttons to give participants more incentive to articulate their intent and choice of service using textual input only. 

The result of our crowdsourcing exercise was that $160$ utterances were added to the baseline training set of $150$ utterances. This resulted in a total of $310$ utterances which are then mapped to lower case prior to use in building the dialog model. 

\begin{figure}[t]
    \includegraphics[width=0.99\linewidth]{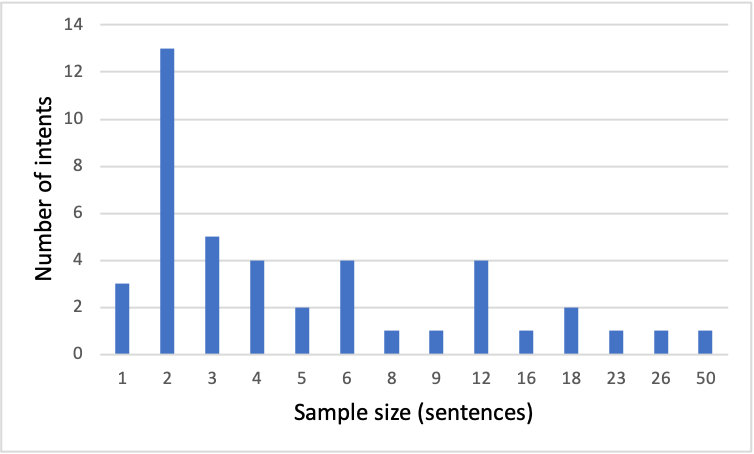}
\caption{Distribution of number of sentences over the intents for \faq and \ff}\label{fig:distribution}
\end{figure}

Figure~\ref{fig:distribution} shows the number of intents (y-axis) by training sample size (x-axis) for both use cases combined. Evidently, the distribution is highly uneven especially for \ff which can lead to a bias in favour of intents with greater numbers of training instances. 
This issue is mitigated and the training data is rebalanced with the use of data generated from the third data source that is user queries received through the firm's website.

\begin{figure}[t]
    \hspace{13.0em}
    \includegraphics[width=0.6\linewidth]{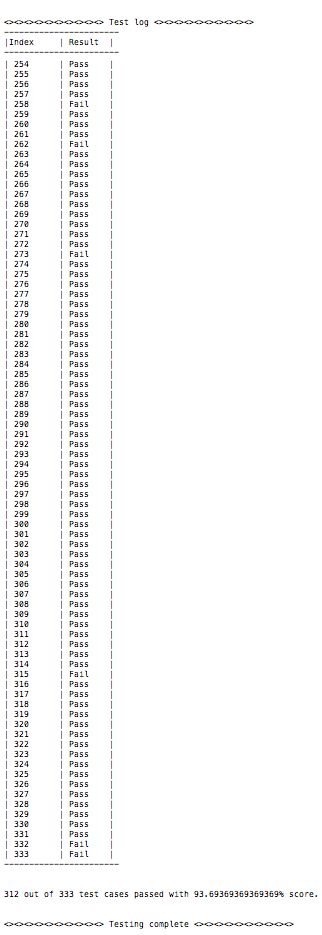}
\caption{Test accuracy on regression test cases}\label{fig:Regression Testcases}
\end{figure}

Although modest in size, the training dataset has proven sufficient to train the dialog model in \lex. The \lex model achieves an accuracy of 93.69 \% when tested on $333$ test sentences. Due to limited space we only show a portion of test cases in the figure~\ref{fig:Regression Testcases}. First column in the figure is the index of the test sentence. Last column shows whether the model has correctly classified the sentence (Pass) or not (Fail), where a class represents an intent in one of the three bots.

\section{Additional features}\label{sec:add features}
A number of additional features and functions have been added to the agent to improve the conversation flow and overall user experience.

\subsubsection{Persist Linguistic context}
First among these is a need to persist linguistic context so that unnecessary repetition or clarification can be avoided. To achieve this, we store information from previous intents and slots. For example, if a user enquires about the price for a service and then changes intent to speak to a solicitor regarding this service, both the previous intent and the previous service are maintained as part of the linguistic context. This allows more seamless switching of intents within a use case or switching from an intent in one use case to an intent in another use case. Storing the previous intent also helps the system interpret follow-up context based conversation, for example:\\

\noindent
\textit{User}: I want to know the price of selling a business.\\
\textit{Sys}: It is very difficult to answer this question unless we have more information regarding your case. We would be happy to ring you to find out more if you can provide us with your contact details. You will not incur any charges until you have accepted any estimate which we give you.\\
\textit{User}: How about an NDA?\\
\textit{Sys}: We can provide a mutual NDA for a fixed price of £175 plus VAT.\\

In this example, the system interprets the ``Price" context by storing the ``\textit{Cost}" intent (or input type) and the service as ``selling a business". For the follow-up question, the system recognises ``How about..." and ``What about..." style sentences and applies the last recorded input type, which in this case is the ``\textit{Cost}", i.e. price for preparing an ``NDA". 



\subsubsection{Restart / Resume}
We give user an option to restart (i.e. start the conversation again) or resume (i.e. pick up from the last conversation). In the latter, the stored linguistic context is used to pick up from where the conversation was left. To facilitate this feature, we define an intent in the parent bot, \textit{``restart\_intent"} that is invoked with sentences such as ``restart the session" and ``Can I restart the session?". For both restart and resume intents, the system prompts a clarification question such as ``Are you sure you want to restart?". To capture user response to this question a slot is defined, \textit{``restart\_slot"}, that can take values such as ``yes", ``sure", ``no" etc.

\subsubsection{Fallback}
It is expected that some of the input utterances will not be identified by the model. In such cases, it is best to avoid generic responses such as ``Sorry I do not understand.". To address this, we define a ``\textit{fallback}" intent in the parent bot. This intent is triggered when the model fails to assign a confidence score of higher than $0.4$ to any intent. A confidence score informs how likely the user utterance belongs to an intent. It's value falls between $0.0$ and $1.0$. Once this intent is triggered, the system suggests the user to enter their contact details so that they may subsequently be contacted by a representative of the firm to assist them with their case. The score of $0.4$ was derived empirically based on multiple trials of the \lex intent classifier. This score indicates that a given utterance is significantly different to the training data, and hence not recognised by the model. We later manually classify the unidentified sentences and include them in the training data to reduce further invocations of the ``\textit{fallback}" intent.

\subsubsection{Automated Testing}
To test the overall bot performance with these additional features followed by a use case (\ff or \faq) or vice versa we have curated various single and multi turn conversations as part of a 'conversation test set'. 'conversation test set' helps to test the system end-to-end. This test set consists of more than $300$ conversations which aims to cover all possible conversation scenarios including switching from one use case or feature (as outlined above) to another. Each scenario shown below consists of multiple conversations consisting of user utterances.
\begin{itemize}
    \item \ff
    \item \faq
    \item Fallback in the parent bot
    \item Restart/ Resume
    \item \ff followed by restart / resume
    \item \faq followed by restart / resume
    \item \ff followed by fallback
    \item \faq followed by fallback
    \item \faq followed by \ff followed by \faq
    \item \ff followed by \faq followed by \ff
    \item \faq followed by fallback in the child bot-\faq
    \item \ff followed by fallback in the child bot-\ff
\end{itemize}

An example for \ff followed by \faq is shown below,\\
\textit{User}: can you help me with employment contract?\\
\textit{Sys}: Are you an employee or an employer?\\
\textit{User}: employee	\\
\textit{Sys}: What is your name?\\
\textit{User}: Jon	\\
\textit{Sys}: What is your phone number?\\
\textit{User}: $07423333333$\\
\textit{Sys}: What is your email address?\\
\textit{User}: jon@xyz.com\\
\textit{Sys}: Do you require assistance with the review of an existing contract or would you like a new contract to be drafted?\\
\textit{User}: i would like review of an existing one.\\
\textit{Sys}: What is your preferred contact method? If by telephone, are there any specific times of the day we should avoid?\\
\textit{User}: email, thanks.\\
\textit{Sys}: Is your matter urgent (i.e. needs to be finalised within the next 48 hours)?
\textit{User}: No\\
\textit{Sys}: Thanks for that. One of our legal experts will contact you as soon as possible. Is there anything else we can help you with?\\
\textit{User}: How long will the process take?\\
\textit{Sys}: This varies from case to case dependant on the complexity of the issue. For straightforward reviews we aim for a turnaround of one or two days.

We set up testing via \aws Command Line Interface (commonly known as \aws cli) that tests the intent identification capability of the model. Each exchange in a conversation is submitted to \lex and once all the conversations are tested, a report is generated that shows the accuracy of the model, that is, the number of utterances in a conversation correctly identified by \lex. This automated testing conducted weekly, in addition to regression testing, also helps in debugging issues and identifies any bugs in the algorithms designed to control the conversation flows.

\subsubsection{Free form user input}
There are multiple ways to fill slots. The bot can be trained to match a keyword (a slot value) in the user text if there is an exclusive list of predefined slot values. Displaying a set of choices (slot values) in the form of a button is another way to fill the slot.
In addition to using these approaches, we allow users to enter free form text for a number of questions asked in \ff case. For instance, the bot asks the user to provide a description of their legal case to fill ``\textit{case\_desc}" slot. It is not possible to predetermine all the responses to this question due to its varied responses. Hence, for questions such as above the user responds in the form of free form text. An option of free form text entry is not provided in \lex so we design a heuristic that assigns a dummy slot value such as ``to\_be\_filled" to the ``\textit{case\_desc}" slot. In the next parse of the algorithm, if slot value is ``to\_be\_filled", algorithm replaces the dummy slot value with the user input.

\subsubsection{Multiple line bot response}
To further improve the user experience we split a bot response consisting of multiple lines into multiple responses. We make use of regular expressions to divide a long response that is returned to the user in the form of multiple responses. We restrict each response with no more than three sentences. For example, in the text shown below, the bot divides the response to return it as a sequence of responses. The first response consists of initial three sentences and the rest of the sentences are sent as a second response to the user. \\

``If you would like advice concerning your contracts I will be happy to help. If you are able to email copies of the contracts you would like advice on please send them to chatbot@xyz.co.uk. If you could provide a few more details and answer the following eight questions, I will get one of our legal experts to contact you to discuss your requirements. You will not incur any charges until you have accepted any estimate which we give you. What is your first name?"\\

\section{Conclusion and future work}\label{conclusion}
In this paper we present a retrieval-based dialog system for the legal profession that supports the answering of FAQs and fact finding by identifying an appropriate legal service and recording case details. We describe the process by which requirements are acquired and the platform selected, along with the architecture and linguistic resources created in developing the system. In conclusion, the model achieved the accuracy of 93.69\% on the regression test set. We also describe the proposed hierarchical bot design with multi-level delegation and additional features we developed to improve user experience.

One limitation of our work is that we heavily rely on the response resource designed with the help of legal experts. In future, we plan to develop a generative model that, unlike retrieval models, would automatically generate responses to the user queries. Generative model introduces variability in bot responses. A further study would be needed to identify resources with multiple sentence paraphrases to train this model. We also plan to extend the bot to other use cases such as booking an appointment.

%
%
%
\bibliographystyle{splncs04}
\bibliography{mybib}

\begin{thebibliography}{10}
\providecommand{\url}[1]{\texttt{#1}}
\providecommand{\urlprefix}{URL }
\providecommand{\doi}[1]{https://doi.org/#1}

\bibitem{bessho2012dialog}
Bessho, F., Harada, T., Kuniyoshi, Y.: Dialog system using real-time
  crowdsourcing and twitter large-scale corpus. In: Proceedings of the 13th
  Annual Meeting of the Special Interest Group on Discourse and Dialogue. pp.
  227--231 (2012)

\bibitem{bordes2016learning}
Bordes, A., Boureau, Y.L., Weston, J.: Learning end-to-end goal-oriented
  dialog. arXiv preprint arXiv:1605.07683  (2016)

\bibitem{chen2017survey}
Chen, H., Liu, X., Yin, D., Tang, J.: A survey on dialogue systems: Recent
  advances and new frontiers. Acm Sigkdd Explorations Newsletter
  \textbf{19}(2),  25--35 (2017)

\bibitem{chen2018deep}
Chen, Y.N., Celikyilmaz, A., Hakkani-Tur, D.: Deep learning for dialogue
  systems. In: Proceedings of the 27th International Conference on
  Computational Linguistics: Tutorial Abstracts. pp. 25--31 (2018)

\bibitem{gao2018neural}
Gao, J., Galley, M., Li, L.: Neural approaches to conversational ai. In: The
  41st International ACM SIGIR Conference on Research \& Development in
  Information Retrieval. pp. 1371--1374 (2018)

\bibitem{serban2015building}
Serban, I.V., Sordoni, A., Bengio, Y., Courville, A., Pineau, J.: Building
  end-to-end dialogue systems using generative hierarchical neural network
  models. arXiv preprint arXiv:1507.04808  (2015)

\bibitem{serban2015survey}
Serban, I.V., Lowe, R., Henderson, P., Charlin, L., Pineau, J.: A survey of
  available corpora for building data-driven dialogue systems. arXiv preprint
  arXiv:1512.05742  (2015)

\bibitem{vinyals2015neural}
Vinyals, O., Le, Q.: A neural conversational model. arXiv preprint
  arXiv:1506.05869  (2015)

\bibitem{yu2016chatbot}
Yu, Z., Xu, Z., Black, A.W., Rudnicky, A.: Chatbot evaluation and database
  expansion via crowdsourcing. In: Proceedings of the chatbot workshop of LREC.
  vol.~63, p.~102 (2016)

\bibitem{zue2000juplter}
Zue, V., Seneff, S., Glass, J.R., Polifroni, J., Pao, C., Hazen, T.J.,
  Hetherington, L.: Juplter: a telephone-based conversational interface for
  weather information. IEEE Transactions on speech and audio processing
  \textbf{8}(1),  85--96 (2000)

\end{thebibliography}
%


\end{document}